\documentclass[twocolumn,aps,pre,longbibliography]{revtex4-2}

\usepackage{tabularx}
\usepackage[latin1]{inputenc}
\usepackage{graphicx}
\usepackage{latexsym}
\usepackage{amssymb}
\usepackage{amsmath}
\usepackage{color}
\usepackage{tikz}
\usepackage{lipsum}
\usepackage{soul}
\usepackage{cancel}
\usepackage{ulem}

\usepackage{amssymb}

\begin{document}
\def\d{{\rm d}}
\def\ex{{\rm e}}
\def\im{{\rm i}}
\def\e{{\bf e}}
\def\E{{\bf E}}
\def\F{{\bf F}}
\def\M{{\bf M}}
\def\J{{\bf J}}
\def\iv{{\rm iv}}
\def\sign{{\rm sign}}
\def\Ecal{\mathcal{E}}
\def\Ncal{\mathcal{N}}
\def\Lcal{\mathcal{L}}
\def\Jcal{\mathcal{J}}
\def\Bcal{\mathcal{B}}
\def\barBcal{{\bar{\mathcal{B}}}}
\def\Ncal{\mathcal{N}}
\def\barNcal{{\bar{\mathcal{N}}}}
\def\Qcal{\mathcal{Q}}
\def\barQcal{{\bar{\mathcal{Q}}}}
\def\smalze{{\scriptscriptstyle{(0)}}}
\def\smalun{{\scriptscriptstyle{(1)}}}
\def\smaln{{\scriptscriptstyle{(n)}}}
\def\smalf{{\scriptscriptstyle{f}}}
\def\smali{{\scriptscriptstyle{i}}}
\def\smalif{{\scriptscriptstyle{if}}}
\def\beq{\begin{eqnarray}}
\def\eeq{\end{eqnarray}}

\newcommand{\stc}[1]{\textcolor{magenta}{\st{#1}}}
\newcommand{\tc}[1]{\textcolor{magenta}{#1}}


\title{Nonanomalous heat transport in a one-dimensional composite chain.}
\author{Piero Olla}
\thanks{Email address for correspondence: olla@dsf.unica.it}
\affiliation{ISAC-CNR and INFN, Sez. Cagliari, I--09042 Monserrato, Italy.}

\begin{abstract}
Translation-invariant low-dimensional systems are known to exhibit anomalous 
heat transport. However, there are systems, such as the coupled-rotor chain, 
where translation invariance is satisfied, yet transport remains diffusive.
It has been argued that the restoration of normal diffusion occurs
due to the impossibility of defining a global stretch variable with
a meaningful dynamics. In this Letter, an alternative mechanism is
proposed, namely, that the transition to anomalous heat transport
can occur at a scale that, under certain circumstances, may diverge 
to infinity. To illustrate the mechanism, I consider the case of a
composite chain that conserves local energy and momentum as well as 
global stretch, and at the same time obeys, in the continuum limit, 
Fourier's law of heat transport.  It is shown analytically that for 
vanishing elasticity the stationary temperature profile 
of the chain is linear; for finite elasticity, the same property 
holds in the continuum limit.

\end{abstract}
\maketitle

\section{Introduction}
\label{Introduction}
Heat transport in solids is described on a phenomenological level by Fourier's law; the
description fails, however, in low-dimensional systems, where
heat transport takes an anomalous character such that the thermal conductivity of the material 
diverges with the sample size 
\cite{lepri03,dhar08,benenti22}.
A standard method for the evaluation of thermal conductivity in solids is provided
by the
Green-Kubo formula \cite{kubo} (see \cite{dhar08} for a simple derivation). In low 
dimensional systems, however, the heat current fluctuation correlation, on which the 
formula is based, diverges at large-scale, which prevents direct application of the method
and suggests breakup of normal transport.
Indeed, analysis of such divergences
by renormalization techniques first allowed researchers to determine the anomalous scaling exponent
for the thermal conductivity in low-dimensional systems \cite{narayan02}
and implied
that a coarse-grained description of the fluctuations in terms of field 
variables in a laboratory reference frame must take into account advection terms analogous 
to those in the Eulerian description of a fluid. 

The analogy in the relation between Lagrangian and Eulerian
description in a fluid, and the dynamics in the continuum limit,
of a low-dimensional solid, was recognized in \cite{chetrite09} and constitutes the basis for the 
derivation of the Nonlinear Fluctuating Hydrodynamic (NFH) theory \cite{spohn14}.
The relevance of the fluid mechanic point of view in the description of heat transport in 
a low-dimensional solid is corroborated by the fact that the same anomalous behaviors
are observed in one-dimensional particle models where the only interaction 
is provided by collisions, and which, at a coarse-grained level, can be described as 
bona fide one-dimensional fluids \cite{kundu16,miron19}. 

The key mechanism leading to the  divergence of the field equations, and hence to anomalous heat 
conduction in the systems under consideration, appears to be the simultaneous conservation locally
of energy and momentum \cite{prosen00,bonetto00,narayan02}. More recently, an additional condition
has been identified in the fact that the global stretch of the system must have
a dynamical content \cite{spohn14a,das14}. If any such condition
is violated---e.g. if the atoms in the chain interact with a substrate, leading to translation 
invariance violation, or if,
as in the case of the coupled-rotor chain \cite{giardina00,gendelman00}, 
it is not possible to define a total 
stretch for the system---normal diffusion 
is recovered. In the same way, systems, such as
the zero-range model \cite{evans05} and the Kipnis-Presutti model \cite{kipnis82} to name a few, 
in which energy is randomly exchanged between atoms without momentum conservation,
are characterized by normal heat conduction.

To date, all analytical models of low-dimensional heat transport are based on mimicking
the role of anharmonicity in spatially redistributing the vibration energy along the chain,
by adding a stochastic component to the dynamics.  The strategy to microscopically
implement stochasticity  is not unique. In \cite{lepri09}, random
collisions are assumed, with pairs of neighboring atoms exchanging
momentum while their total energy remains constant. In other models, 
three-atom interactions are required to accommodate
the joint conditions of energy and momentum conservation. In \cite{basile06}, the stochastic 
component of the dynamics is realized by a random walk in momentum space on the
constant energy surface of the system. 
NFH predicts that for generic interaction potentials, the large-scale dynamics of 
energy and momentum preserving one-dimensional chains should
fall in the universality class of the Kardar-Parisi-Zhang model 
\cite{spohn14}.
There are special cases, however,  in which the predictions of the NFH theory do not apply
\cite{spohn14,lee-dadswell15}, with finite-size effects, as well as weak chaos in the interaction,
making the detection of universal behaviors difficult \cite{lepri20}.

The situation as regards experiments and numerical (atomistic) simulation of more
realistic systems is equally complicated. 
Results are indeed often dependent on the properties of
the material and the experimental or numerical technique adopted (see \cite{benenti22} and 
references therein for an extended discussion). Of particular interest
is the possible presence of a diffusive range at 
small scales, complicating 
the measurement of the anomalous scaling exponents predicted by the theory;
such crossover behaviors are indeed predicted in particle systems \cite{miron19}.

The purpose of the present Letter is to study the crossover from small-scale thermal diffusion
to large-scale anomalous heat conduction 
in the specific example of a ``composite'' chain, in which 
atoms interact with their neighbors through harmonic forces and inclusions acting
as random sources and sinks of kinetic energy. Composite materials such as e.g. 
semiconductor perovskites find application in photovoltaics, and proper characterization 
of their thermal properties is particularly important \cite{caddeo16}. 
The total momentum and energy of 
the atoms and the inclusion involved in an interaction are conserved, the ends
of the chain are fixed, and thus
all the conditions for anomalous heat conduction in the system are satisfied.
Yet, the analysis that follows shows that the range in which heat transport is diffusive 
can greatly exceed the range in which the dynamics of the chain is viscous. In particular,
heat transport becomes diffusive at all scales in the continuum limit.

\label{Outline of the model}
\section{Outline of the model}
\begin{figure}
\begin{center}
\includegraphics[draft=false,width=\linewidth]{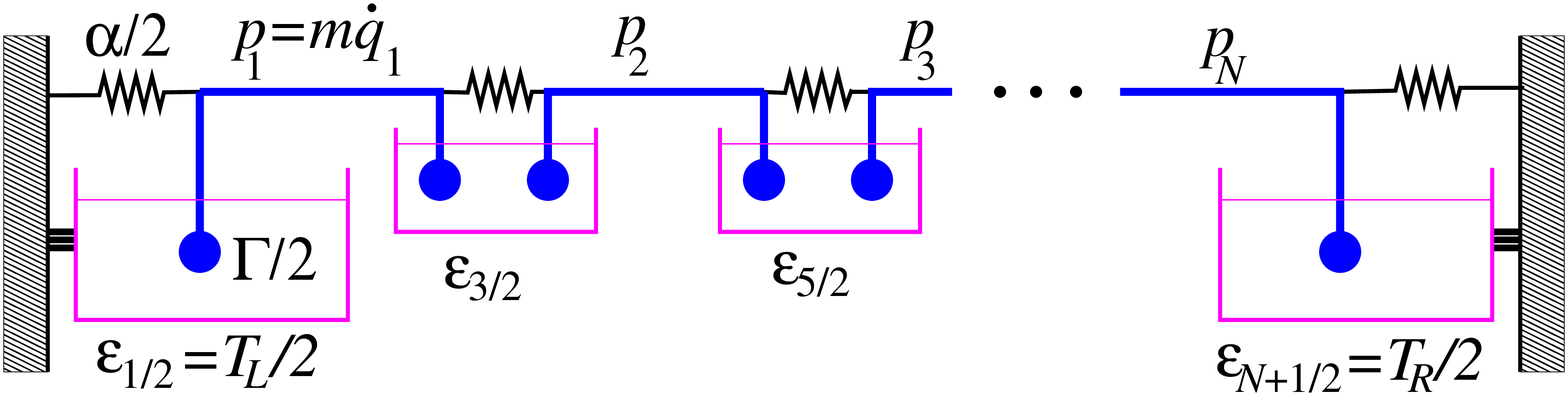}
\caption
{
Sketch of the composite chain: the two-bead assemblies (blue online) represent the individual 
atoms; the cells underneath (magenta online) represent the zero-mass mobile inclusions
and the thermal baths.
}
\label{chain}
\end{center}
\end{figure}

The geometry of the system is illustrated in Fig. \ref{chain}.
Each two-bead assembly represents an atom, with the bars joining the
beads assumed rigid. The cells in the middle represent the inclusions,
which for simplicity are taken to be massless; the cells at the extremes of the chain 
are the heat baths, whose position is fixed. Indicate with $N$ the number of atoms in the
chain and with $L=N\eta$ the chain's length.
The inclusions act on the beads as Langevin 
baths with friction coefficient $\Gamma/2$ and noise amplitude 
\beq
\langle\xi_{k+1/2}(0)\xi_{j+1/2}(t)\rangle=2m\Gamma\langle\epsilon\rangle_{k+1/2}
\delta_{kj}\delta(t).
\label{xixi}
\eeq
The adopted Langevin dynamics may be interpreted as the result
of coarse graining the fast internal degrees of freedom in the inclusions,
with $\epsilon_{k+1/2}$ an energy variable
\footnote{In the case of an approximately linear internal dynamics, 
$\epsilon_{k+1/2}$ would be the energy per degree of freedom of the inclusion.
}
 that is going to be determined 
dynamically from the condition of local energy conservation (see below).

The thermal baths at the chain extremes 
act on the respective atoms in the same way as the inclusions, 
with energy variables 
$\epsilon_{1/2}$ and $\epsilon_{N+1/2}$ replaced in Eq. (\ref{xixi})
by fixed temperatures
$T_L/2$ and $T_R/2$, $T_L-T_R=2\Delta T$
(the Boltzmann constant $k_B$ is set equal to 1 throughout the calculation). 

Indicate with $q_k$ the displacement of the atoms from their equilibrium 
position and with $p_k=m\dot q_k$ the associated momentum. 
As illustrated in Fig. \ref{chain}, the elastic force acts in parallel with that
by the inclusion;
the chain dynamics is then described by the system of equations, in the bulk $1<k<N$,
\beq
\dot p_k&=&\Gamma(p_{k+1}+p_{k-1}-2p_k)/2
\nonumber
\\
 &+&\alpha(q_{k+1}+q_{k-1}-2q_k)/2+\xi_{k-1/2}-\xi_{k+1/2},
\label{bulk}
\eeq 
while at the ends of the chain,
\beq
\dot p_1&=&\Gamma(p_2-2p_1)/2
\nonumber
\\
&+&\alpha(q_2-2q_1)/2+\xi_{1/2}-\xi_{3/2},
\label{left}
\\
\dot p_N&=&\Gamma(p_{N-1}-2p_N)/2
\nonumber
\\
&+&\alpha(q_{N-1}-2q_N)/2+\xi_{N-1/2}-\xi_{N+1/2}
\label{right}
\eeq
(It\^o's prescription is assumed throughout the Letter).
The conservative nature of the noise in Eqs. (\ref{bulk}-\ref{right}) should be noted,
which distinguishes the present model from ones in which local heat baths force the
dynamics, such as e.g. \cite{bolsterli70,bonetto04}. 

It is possible to identify a microscopic elastic timescale $\omega_\eta^{-1}=\sqrt{m/\alpha}$,
with the magnitude of the ratio
\beq
r=\omega_\eta/\Gamma
\label{ratio}
\eeq
determining whether the microscopic dynamics is dominated by elasticity or by the effective
friction generated by the inclusions.
We can take
the continuum limit of Eq. (\ref{bulk}), and the result is 
 \footnote{Note that $x$ in Eqs.
(\ref{wave equation}) and (\ref{noise}) is a Lagrangian variable; note also that 
the equations are linear,
which implies that neither viscosity nor sound speed renormalization is required in the analysis 
that follows.}:
\beq
\partial_t^2 q&=&\partial_x^2(c_s^2+\nu\partial_t)q+\partial_x\xi,
\label{wave equation}
\\
\langle\xi(x,t)\xi(0,0)&=&2(\eta\langle\epsilon\rangle\nu/m)\delta(x)\delta(t),
\label{noise}
\eeq
which describes wave propagation in a viscoelastic (Kelvin-Voigt) medium with sound speed
and viscosity, respectively,
\beq
c_s=\eta\omega_\eta\quad{\rm and}\quad\nu=\eta^2\Gamma.
\label{c_s mu}
\eeq
From here, it is possible to define a viscous scale
\beq
l_\nu=\nu/c_s=\eta/r,
\label{l_nu}
\eeq
which identifies the upper limit of the viscosity-dominated range for the dynamics.

To study the fluctuation dynamics, one needs an equation for the  energy variable
$\epsilon_{k+1/2}$. One obtains such equation by imposing energy conservation in 
the interaction between atoms and inclusions.
The energy budget in the interaction between atoms $k$ and $k+1$, and inclusion $k+1/2$
is obtained, for $1<k<N$,
by evaluating the contribution to the variation of kinetic energy of the two atoms,
$K_{k+1/2}=(p_{k+1}-p_k)^2/(4m):=p_{k+1/2}^2/(4m)$, 
from the forcing by the inclusion. One can write in 
general
\beq
\dot K_{k+1/2}=-\dot E_{k+1/2}+\ldots,
\label{dot K}
\eeq
where $E_{k+1/2}=E(\epsilon_{k+1/2})$ is the internal energy of the inclusion
and the dots stand for 
the contributions from the elastic forces and the neighboring inclusions.
Substituting Eq. (\ref{bulk}) in the left hand side of Eq. (\ref{dot K})
yields then
\beq
\dot E_{k+1/2}&=&2\Gamma\Big(\frac{p^2_{k+1/2}}{4m}-\epsilon_{k+1/2}\Big) 
-\frac{p_{k+1/2}\xi_{k+1/2}}{m},
\label{dot E}
\eeq
where one recognizes in the term $\Gamma p_{k+1/2}^2/(2m)$ the work by the friction forces, 
and in $2\Gamma\epsilon_{k+1/2}$ the average energy provided to the two atoms 
by the fluctuating force.

\label{The purely viscous chain}
\section{The purely viscous chain}
In 
the $\alpha\to 0$ limit, the system provides an example of heat transport by Brownian motion.
Exactly as in the elastic case \cite{rieder67}, it is possible to evaluate the
heat flow from the dynamics of the one-time correlations \footnote{ 
Note that once a constitutive relation $E_{k+1/2}=E(\epsilon_{k+1/2})$ is selected, 
Eqs. (\ref{xixi}-\ref{right}) and (\ref{dot E}), together with the kinematic
condition $p_k=m\dot q$, constitute a system of stochastic differential equation with 
multiplicative noise. Linearity of the stochastic equations, nevertheless, makes the equations
for the correlations analytically solvable.}. Define
\beq
\Pi_{ik}=\langle p_ip_k\rangle,
\ 
Z_{ik}=\langle q_ip_k\rangle,
\
Q_{ik}=\langle q_iq_k\rangle.
\eeq
For $\alpha=0$, $q_i$ is an irrelevant variable, and the equation for $\Pi$ and those 
for $Z$ and $Q$ decouple. The last two variables can then, for the moment, be disregarded.

Let us focus first on the bulk;
from Eq. (\ref{bulk}), one obtains the equation for $\Pi_{ik}$
\beq
2\Gamma^{-1}\dot\Pi_{k,k+l}&=&\Pi_{k+1,k+l}+\Pi_{k-1,k+l}-2\Pi_{k,k+l}
\nonumber
\\
&+&\Pi_{k,k+1+l}+\Pi_{k,k-1+l}-2\Pi_{k,k+l}
\nonumber
\\
&+&4m[(\delta_{l0}-\delta_{l1})\langle\epsilon\rangle_{k+1/2}
\nonumber
\\
&+&
(\delta_{l0}-\delta_{l,-1})\langle\epsilon\rangle_{k-1/2}],
\label{Pi equation 0}
\eeq
where $1<k,k+l<N$. 
At stationarity, one gets from Eq. (\ref{dot E}) 
\beq
4m\langle\epsilon\rangle_{k+1/2}&=&\langle p_{k+1/2}^2\rangle
\nonumber
\\
&=&\Pi_{kk}+\Pi_{k+1,k+1}-2\Pi_{k,k+1},
\label{stationarity}
\eeq
which, substituted into Eq. (\ref{Pi equation 0}), yields
\beq
-2\Pi_{kk}+\Pi_{k+1,k+1}+\Pi_{k-1,k-1}&=&0,
\label{kk}
\\
-2\Pi_{k,k+1}+\Pi_{k,k+2}+\Pi_{k+1,k-1}&=&0,
\label{k,k+1}
\\
\Pi_{k+1,k+l}+\Pi_{k-1,k+l}+\Pi_{k,k+1+l}& &
\nonumber
\\
\qquad+\Pi_{k,k-1+l}-4\Pi_{k,k+l}&=&0
,\quad |l|>1.
\label{k,k+l}
\label{Pi equation}
\eeq
The same procedure can be carried out at $k=1$,
\beq
\Pi_{22}-3\Pi_{11}+2mT_L&=&0,
\label{11}
\\
\Pi_{13}-2\Pi_{12}&=&0,
\label{12}
\\
\Pi_{2l}-4\Pi_{1l}+\Pi_{1,l-1}+\Pi_{1,l+1}&=&0,
\quad l>2,
\label{1l}
\eeq
and
a similar set of equations is produced at $k=N$. 
The system of equations (\ref{kk}-\ref{1l}) has the remarkable property 
that
Eqs. (\ref{k,k+1},\ref{k,k+l},\ref{12},\ref{1l}), which involve out-of-diagonal terms,
decouple from Eqs. (\ref{kk}) and (\ref{11}) on the diagonal. 
Equations (\ref{kk}) and (\ref{11}) tell us that $\Pi_{kk}$ has a linear profile:
\beq
2(mT_L-\Pi_{11})=\Pi_{kk}-\Pi_{k+1,k+1}
\nonumber
\\
=2(\Pi_{NN}-mT_R)=2m\Delta T/N.
\label{linear}
\eeq
On the other hand, Eqs. (\ref{k,k+1},\ref{k,k+l},\ref{12},\ref{1l}) admit the zero solution 
$\Pi_{kl}=0$, $k\ne l$, which is also necessarily unique, since a nonzero
solution could have arbitrary amplitude and lead to negative $\langle (p_l+p_k)^2\rangle$).
From Eq. (\ref{linear}), it is then possible to write \footnote{
The same result could
be obtained by imposing a linear profile for $\langle\epsilon\rangle_{k+1/2}$,
rather than solving self-consistently Eq. (\ref{dot E}). The operation would be
equivalent to replacing the inclusions with zero-mass thermal baths, and would
result in the modification of the the higher-order correlations and
time-dependent component of the statistics.}
\beq
\Pi_{kl}=m\Big(T_L+\frac{2k-1}{N}\Delta T\Big)\delta_{kl};
\label{diagonal}
\eeq 
The temperature profile along the purely viscous chain, $T_k\equiv \Pi_{kk}/m
\simeq 2\langle\epsilon_{k+1/2}\rangle$, 
is thus linear.

For $\alpha=0$, the heat transfer along the chain is mediated by the work on the atoms
by the inclusions;
the average work by inclusion $k-1/2$ on the atom to its right thus coincides with the
heat flux at site $k$
\beq
J_k&=&\Gamma\Big[\langle\epsilon\rangle_{k-1/2}
-\frac{\langle p_k(p_k-p_{k-1})\rangle}{2m}\Big],
\label{J_k}
\eeq
where it is understood that $p_0=0$.
At stationarity, from Eqs. (\ref{stationarity}) and (\ref{diagonal}),
\beq
\frac{J_k}{\Gamma}=\frac{T_L}{2}-\frac{\Pi_{11}}{2m}=\frac{\Pi_{k-1,k-1}-\Pi_{kk}}{4m}
=\frac{\Pi_{NN}}{2m}-\frac{T_R}{2},
\nonumber
\eeq
which, using again Eq. (\ref{diagonal}), implies Fourier's law; after reinstating 
Boltzmann's constant,
\beq
J=\kappa\frac{T_L-T_R}{L},
\qquad 
\kappa=\frac{k_B\nu}{4\eta}.
\label{Fourier}
\eeq
Setting $T_L=T_R=T$, it is easy to verify 
from Eqs. (\ref{stationarity}), (\ref{J_k}) and (\ref{diagonal}),
that at equilibrium $J_k=0$ and equipartition holds: $\langle\epsilon\rangle_{k+1/2}
=\Pi_{kk}/(2m)=T/2$.

\section{The effect of finite elasticity}
For 
finite $\alpha$, part of the heat transfer is mediated by the elastic forces, with a
contribution to the heat flux \cite{lepri03}:
\beq
\delta J^{el}_k=\alpha(Z_{kk}-Z_{k+1,k})/(2m).
\label{J^el(Z)}
\eeq
To evaluate $\delta J^{el}_k$, we need an equation for $Z_{ik}$.
Indicate
\beq
A_{ik}=\Gamma Z_{ik}+\alpha Q_{ik}.
\eeq
The stationarity conditions $\dot Z_{k,k+l}+\dot Z_{k+l,k}=0$ and 
$\dot Z_{k,k+l}-\dot Z_{k+l,k}=0$ take the form, in the bulk, from Eq. (\ref{bulk}):
\beq
-4\Pi_{k+l,k}/m=A_{k+l,k+1}+A_{k+l,k-1}
\nonumber
\\
-2A_{k+l,k}+A_{k,k+l+1}+A_{k,k+l-1}-2A_{k,k+l},
\label{eq1}
\\
A_{k+l,k+1}+A_{k+l,k-1}-2A_{k+l,k}
\nonumber
\\
=A_{k,k+l+1}+A_{k,k+l-1}-2A_{k,k+l}.
\label{eq2}
\eeq
Equations (\ref{eq1}) and (\ref{eq2}) imply
$-2\Pi_{k+l,k}/m=A_{k+l,k+1}+A_{k+l,k-1}-2A_{k+l,k}$, better rewritten as
\beq
\partial_k^2A_{lk}=-2\Pi_{lk}/m, \quad
1<k,l<N.
\label{previous}
\eeq
where $\partial_k$ indicates finite difference: $\partial_k^2f=f_{k+1}+f_{k-1}-2f_k$.
Identify with a bar the zero-viscosity
component of quantities and with a tilde the respective correction,
$\Pi=\bar \Pi+\tilde \Pi$, and make the ansatz (to be verified a posteriori) 
that if the chain is sufficiently short viscosity will dominate thermal fluctuations. 
It is then possible to set in Eq.
(\ref{previous}) $\Pi_{lk}\simeq\bar\Pi_{lk}=\bar\Pi_{ll}\delta_{lk}$ [see Eq. (\ref{diagonal})],
which yields the solution
\beq
A_{lk}\simeq a_l+ b_lk-\bar\Pi_{ll}|k-l|/m.
\label{lowest A}
\eeq
To determine the coefficients $a_l$ and $b_l$, 
one needs boundary conditions, which are
provided by imposing stationarity at the ends of the chain, $\dot Z_{1k}=\dot Z_{Nk}=0$;
exploiting Eqs. (\ref{bulk}), (\ref{left}) and (\ref{right}):
\beq
2\Pi_{l1}/m&=&2A_{l1}-A_{l2},
\nonumber
\\ 2\Pi_{lN}/m&=&2A_{lN}-A_{l,N-1},\quad
1<l<N.
\label{B.C.}
\eeq
From
$\Pi_{1l}\simeq\Pi_{lN}\simeq 0$, $1<l<N$,
the following relations are then obtained:
\beq
A_{l2}\simeq 2A_{l1},\quad
A_{l,N-1}\simeq 2A_{lN},\quad 1<l<N.
\label{B.C. leftright}
\eeq
Substituting Eq. (\ref{lowest A}) into Eq. 
(\ref{B.C. leftright})
yields 
\beq
a_l\simeq\frac{l\bar\Pi_{ll}}{m},
\quad 
b_l\simeq\frac{\bar\Pi_{ll}}{m}\frac{N+1-2l}{N+1},\quad 1<l<N.
\eeq
Exploiting Eq. (\ref{lowest A}) and the relation 
$\bar\Pi_{kk}-\bar\Pi_{k+1,k+1}=2m\Delta T/N$  [see Eq. (\ref{linear})],
we obtain the expression, valid for $1<l\le k<N$,
\beq
Z_{lk}&\simeq&\frac{A_{lk}-A_{kl}}{2\Gamma}=\frac{l(N+1-k)(\bar\Pi_{ll}
-\bar\Pi_{kk})}{m\Gamma(N+1)}
\nonumber
\\
&=&\frac{2l(N+1-k)(l-k)\Delta T}{\Gamma N(N+1)}.
\label{Z}
\eeq
Substituting Eq. (\ref{Z}) into Eq. (\ref{J^el(Z)}) finally yields
\beq
\delta J_k^{el}\simeq-\frac{\alpha(1+k)(N+1-k)\Delta T}{\Gamma mN(N+1)},
\label{J^el}
\eeq
and hence, by comparing with Eq. (\ref{Fourier}),  the estimate
\beq
\delta
J^{el}/J\sim r^2N=\eta L/l_\nu^2,
\label{estimate}
\eeq
which tells us that as long as
\beq 
L\ll l_\kappa=l^2_\nu/\eta,
\label{l_kappa}
\eeq 
heat transport remains diffusive.
A similar estimate holds for the momentum fluctuation amplitude,
$\tilde\Pi/\bar\Pi\sim r^2N$ (see Supplemental Material \cite{supp}), which confirms the ansatz
at the basis of Eq. (\ref{lowest A}).
Note that in the continuum limit
$\eta\to 0$ (all macroscopic quantities $L$, $Nk_BT$, $Nm$, $\nu$ and $c_s$
fixed and finite), 
the diffusive scale $l_\kappa$ goes to infinity and
Fourier's law holds irrespective of the sample size.

 An interesting question concerns
the role of possible violations of stretch conservation in the regime $L\ll l_\kappa$.
By construction, for finite $\alpha$, the total stretch $\delta L$ is controlled 
by elasticity, which means that, strictly speaking, the
total stretch is conserved. One may nevertheless argue that if, for
$l_\nu\ll L\ll l_\kappa$ (that is the range where the dynamics of the chain is elastic), 
the ratio $\delta L/L$ in the continuum limit were to diverge to infinity,
the chain would be behaving as if its endpoints were unconstrained. In other words, 
diffusive transport for $l_\nu\ll L\ll l_\kappa$ could be the consequence of insufficient
conservation of stretch.  We show below that this is not the case.

The continuum limit $\eta\to 0$ ($L$, $Nk_BT$, $Nm$, $\nu$ and $c_s$ fixed and finite) 
corresponds to a regime 
$L\ll l_\kappa$ such that Eq. (\ref{diagonal})
applies. It is then possible to estimate for the stretch
$\delta L\sim \sqrt{N\langle q_{k+1/2}^2\rangle}$, where 
$q_{k+1/2}=q_{k+1}-q_k$; 
for small deviations from 
equilibrium, $T_L\simeq T_R\sim T$, 
$\langle q_{k+1/2}^2\rangle\sim v_{th}^2m/\alpha=(v_{th}/\omega_\eta)^2$,
where $v_{th}=\sqrt{k_BT/m}$ is the thermal velocity, which remains finite in the limit.
From Eqs. (\ref{c_s mu}) and (\ref{l_nu}) one then gets
\beq
\frac{\delta L}{L}\sim\frac{N^{1/2}v_{th}}{L\omega_\eta}=\frac{N^{1/2}\eta v_{th}}{Lc_s}
=\frac{\sqrt{l_\nu\eta}}{L}\frac{v_{th}}{c_s},
\label{stretch}
\eeq
which tells us that the relative stretch vanishes in the continuum limit (all quantities in the
right hand side of the formula remain finite, except $\eta$, which vanishes); this suggests
that stretch conservation violations do not play a role in the dynamics under consideration.
(An alternative derivation of
the result is provided in the Supplemental Material \cite{supp}.) 

\section{Conclusion}
The present analysis shows
that heat transport in a composite chain with 
a purely viscous microscopic dynamics, obeys Fourier's law.
By construction, no internal forces are generated in response to 
stretching of the system, and we thus have another example, beside that of the 
coupled-rotor chain, of a system conserving local energy and momentum, in which
global stretch is not conserved and heat transport is diffusive.

If elasticity is finite, but there is a viscous 
range extending to macroscopic scales, the range of scales where heat transport is diffusive 
greatly exceeds the viscous range, and extends to infinity in the continuum limit. This is
the main result of the Letter.  The global 
stretch fluctuations vanish in the limit, which means that diffusive heat transport
in such composite chain,
contrary to the case of the coupled-rotor chain, cannot be explained by
non-conservation of the stretch. 

Some questions remain open. Is anomalous heat transport in the composite chain
recovered at scales larger than the diffusion length $l_\kappa$? 
Is the existence of an extended diffusive range 
a common property of viscoelastic one-dimensional chains? The answer is probably yes in both
cases \footnote{Note added after publication: Eqs. (15-17) and Eq. (29) in the
present Letter coincide with Eqs. (14-15) and Eq. (36) in Ref. \cite{lepri09};
the answer to the first question is therefore yes; the answer to the second question
is again yes in the case of the model of Ref. \cite{lepri09}.
}, but, to prove the statement, going beyond the present perturbative---model-dependent 
analysis, would be required.

\bibliography{sample}

\onecolumngrid
\newpage
\centerline{\bf SUPPLEMENTAL MATERIAL}
\vskip 10pt

\centerline{\bf A: Finite elasticity correction to the momentum fluctuation amplitude}
\vskip 5pt
An equation for the stationary momentum correlation in the bulk, valid for generic
$\alpha$, can be obtained from Eqs. (2) and (11):
\beq
\partial_k^2\Pi_{kk}+(\alpha/\Gamma)(Z_{k+1,k}+Z_{k-1,k})=0.
\label{Pikk0}
\eeq
By exploiting Eq. (34), Eq. (\ref{Pikk0}) takes the form, valid for small $r$
and large $N$,
\beq
\partial_k^2\Pi_{kk}\simeq\frac{4\alpha\Delta T(N/2-k)}{\Gamma^2N^2}.
\label{Pikk}
\eeq
Equation (\ref{Pikk}) has the general solution 
\beq
\Pi_{kk}\simeq\frac{8\alpha\Delta T}{3\Gamma^2N^2}(N/2-k)^3+c_0+c_1k.
\eeq
The homogeneous part of the solution is fixed by the boundary conditions
\beq
\Pi_{11}=mT_L,
\qquad
\Pi_{NN}=mT_R.
\eeq
From Eq. (22) one obtains then the expression for the elastic correction
\beq
\tilde\Pi_{kk}=\Pi_{kk}-\bar\Pi_{kk}\simeq
\frac{8\alpha\Delta T}{3\Gamma^2N^2}[(N/2-k)^3-N^2(N/2-k)/4].
\eeq
and hence the estimate [see Eq. (36)]
\beq
\frac{\tilde\Pi_{kk}}{m\Delta T}\sim r^2N\sim\frac{\delta J^{el}_k}{J}.
\eeq
\vskip 15pt
\centerline{\bf B: Stretch fluctuation amplitude --- alternative calculation.}
\vskip 5pt
The stretch fluctuation can be estimated as
$\delta L\sim \langle q_j^2\rangle^{1/2}$, where $\langle q_j^2\rangle\equiv Q_{jj}$.
By switching to Fourier
space in both space and time, we easily obtain from Eqs. (6) and (7) the fluctuation spectrum
\beq
Q_{k\omega}=\frac{2\nu\epsilon k^2}{\rho[(\omega^2-c_s^2k^2)^2+\nu^2k^4]}
=\frac{\nu k_BT k^2}{\rho[(\omega^2-c_s^2k^2)^2+\nu^2k^4]},
\eeq
where $\rho=m/\eta$ is the chain density and 
$k$ indicates here wavenumbers. The denominator of $Q_{k\omega}$ has roots
\beq
&&kl_\nu\gg 1:\qquad \omega_1\simeq \pm\im\nu k^2,\quad\omega_2=\pm\im c_s^2/\nu,
\nonumber
\\
&&kl_\nu\ll 1:\qquad \omega_1\simeq c_sk\pm\im\nu k^2,\quad\omega_2=-c_sk\pm\im\nu k^2;
\nonumber
\eeq
we thus get for $kl_\nu\gg 1$
\beq
Q_k=
\int\frac{\d\omega}{2\pi}\frac{\nu k_BTk^2}{\rho(\omega^2+\nu^2k^4)(\omega^2+c_s^4/\nu^2)}
=\frac{\nu k_BT}{2\rho c_s^2(c_s^2/\nu+\nu k^2)}
\simeq \frac{k_BT}{2\rho c_s^2k^2},
\label{Q_k1}
\eeq
while for $kl_\nu\ll 1$ we have
\beq
Q_k&=&
\int\frac{\d\omega}{2\pi}\frac{\nu k_BTk^2}{\rho
[(\omega-c_sk)^2+\nu^2k^4][(\omega+c_sk)^2+\nu^2k^4]}
\nonumber
\\
&=&\frac{k_BT}{4\rho\nu(\nu^2k^4+c_s^2k^2)}
\simeq \frac{k_BT}{4\rho c_s^2k^2}.
\label{Q_k2}
\eeq
From here we get the displacement fluctuation amplitude in a chain of length $L$ 
\beq
\langle q^2\rangle\sim\int_{kL>1}\d k Q_k\sim \frac{k_BTL}{\rho c_s^2}\sim (v_{th}/c_s)^2\eta L,
\eeq
and hence the estimate for the stretch, valid for both $l_\nu\ll L$ and $l_\nu\gg L$,
\beq
\delta L\sim \langle q^2\rangle^{1/2}\sim\frac{v_{th}\sqrt{\eta L}}{c_s}=
\frac{v_{th}N^{1/2}\eta}{c_s}
\eeq
[compare with Eq. (38)].

\end{document}